\documentclass[aps,prb,twocolumn,showpacs,floatfix,superscriptaddress]{revtex4-2}

\usepackage{bm}
\usepackage{float}
\usepackage{bbding}
\usepackage{graphicx}
\usepackage[dvipsnames]{color}
\usepackage{amsfonts}
\usepackage[figuresright]{rotating}
\usepackage{amssymb}
\usepackage{amsmath}
\usepackage{pifont}
\usepackage{psfrag}
\usepackage{subfigure}
\usepackage{multirow}
\usepackage{tabularx}
\usepackage{textcomp}
\usepackage{units}
\usepackage{hyperref}
\usepackage{comment}
\usepackage{longtable}
\hypersetup{
 pdfnewwindow=true, colorlinks=true,
 linkcolor=blue, anchorcolor=blue,
 citecolor=blue, filecolor=blue,
 menucolor=blue, urlcolor=blue}

\def\nn{\nonumber}

\def\beq{\begin{eqnarray}}
\def\eeq{\end{eqnarray}}

\renewcommand{\v}[1]{\ensuremath{\mathbf{#1}}} 

\let\baraccent=\= 
\renewcommand{\=}[1]{\stackrel{#1}{=}} 

\definecolor{light-blue}{rgb}{0.8,0.85,1}
\definecolor{cyan}{rgb}{0,.5,.9}
\definecolor{violet}{rgb}{.7,0,.9}

\makeatletter

\begin{document}

\title{Nonlinear photomagnetization in insulators}

\author{Bernardo S.\ \surname{Mendoza}}
\affiliation{Centro de Investigaciones en \'Optica, A.C., Le\'on, Guanajuato 37150, Mexico}
\affiliation{Max Planck Institute for the Structure and Dynamics of Matter, Luruper Chaussee 149, 22761 Hamburg, Germany}

\author{Norberto\ \surname{Arzate-Plata}}
\affiliation{Centro de Investigaciones en \'Optica, A.C., Le\'on, Guanajuato 37150, Mexico}

\author{Nicolas\ \surname{Tancogne-Dejean}}
\affiliation{Max Planck Institute for the Structure and Dynamics of Matter, Luruper Chaussee 149, 22761 Hamburg, Germany}

\author{Benjamin\ \surname{M. Fregoso}}
\affiliation{Department of Physics, Kent State University, Kent, Ohio 44242, USA}

\begin{abstract}
Nonlinear photomagnetization is a process by which an oscillating electric field induces a static magnetization. We show that all 32 crystallographic point groups admit such spin polarization using circularly polarized electric fields to second order (as in the usual spin orientation or inverse Faraday effect) but only 29 points groups admit spin polarization using linearly polarized electric fields to second order. The excluded point groups are the highly symmetric $m$-3$m$, -$43m$ and 432. Using density functional theory we compute the spectrum of the second-order electric spin susceptibility of prototypical semiconductors Te, Se, SnS$_2$, GaAs, InSb and Si which corresponds to nonmagnetic materials with and without inversion symmetry. We show that nonlinear photomagnetization can be  comparable to those of naturally occurring ferromagnets.
\end{abstract}

\maketitle

\section{Introduction}
Control over material's magnetization finds important technological applications in data storage, memory reading/writing, and quantum information\cite{Zutic2004,Dresselhaus1955,Bychkov1984,Aronov1989,Edelstein1990,Pitaevskii1961,Meier1984,Driel2000}. Materials chosen for applications usually do not have inversion or time-reversal symmetry because the ground state of such materials has spin-degenerate bands\cite{Zhang2014,Riley2014,Gotlieb2018}. This limitation is partially lifted in the nonlinear regime where it is possible to generate a macroscopic spin polarization using electric fields\cite{Aronov1989,Edelstein1990,Pitaevskii1961,Meier1984,Driel2000,Kimel2005,Tarasenko2005,Pesin2012,Stanciu2007,Alebrand2012,Lambert2014,Mangin2014,Kirilyuk2010,Tokman2020,Gao2020,Tanaka2020,Banerjee2022,Berritta2016,Scheid2019,Freimuth2016,Cheng2020,Gu2010,Hurst2018,Smolyaninov2005,Battiato2014,Hertel2006,Nadarajah2017,SinhaRoy2020,Wagniere1989,Volkov2002,Mishra2023}. In this is case one is confined to use circularly polarized electric fields (CPE) since the usual/strongest mechanism is the transfer of angular momentum from external sources to electron's spin \cite{Meier1984,Driel2000}. For a review of the recent progress, see Johansson~\cite{Johansson2024}. A linearly polarized electric field (LPE), on the other hand, is not expected to induce a net spin polarization since it carries no angular momentum. Here, we show rigorously that \textit{all} materials can be spin polarized by CPEs to second order (as expected) but a large class of materials also admit spin polarization by LPE as long as the underlying crystal structure is not too symmetric.

The key idea is that linear polarization modifies electron's orbital motion which in turn modifies electron's spin, provided there is spin-orbit coupling (SOC)~\cite{Fregoso2022}. A net spin polarization requires quantum coherence and energy absorption. Here, we uncover another requirement, that the crystal point group must not be $m$-3$m$, -$43m$ or 432 because otherwise the spin density vanishes by symmetry. Hence LPE spin polarization arises from a specific set of orbital motions within a crystal. Where does the spin angular momentum comes from? the net spin density is transferred from other degrees of freedom through internal torques\cite{Atencia2024}. This should be compared and contrasted with the usual spin orientation effects which obviate the need for magnetic materials or SOC because they use CPEs. To demonstrate real-life applications of our results we use large-scale Density Functional Theory (DFT) methods to numerically compute the spectrum of the spin response (pseudo)tensor of \textit{nonmagnetic}, \textit{centrosymmetric} semiconductors Si and SnS$_2$ and compare with the spin response of semiconductors with no inversion symmetry, GaAs, Te, and InSb, and which are expected to exhibit the usual spin orientation effects or inverse Faraday effect. Our work highlights a control knob, the light polarization direction, on the magnetic properties of materials.

\section{Symmetry constrains}
Consider the expansion of the spin polarization in powers of an electric field to second order. More generally, consider the expansion of a macroscopic (time-independent) \textit{pseudovector} $\pmb{\bar{\mathcal{V}}}$ in powers of a (time-dependent) vector $\v{V}$ (schematically),
\begin{align}
\bar{\mathcal{V}}^{a}(\omega) = \zeta^{abc}(\omega) V^{b}(\omega) V^{c}(-\omega),
\label{eq:S_gen}
\end{align}
where $\zeta$ is a third-rank pseudotensor and $\v{V}(\omega)$ is the Fourier component of a homogeneous vector given by 
\begin{align}
\v{V} = \v{V}(\omega) e^{-i \omega t} + c.c.
\label{eq:e-field}
\end{align}
For clarity we omit all frequency dependance in subsequent equations. Summation over repeated indices is implied. In metals the leading contribution to spin polarization is linear, i.e.,  the Edelstein effect~\cite{Johansson2024}, but in insulators Eq.~(\ref{eq:S_gen}) is the leading contribution~\cite{Fregoso2022}.

A transformation of the coordinate frame by an element $M$ of the crystallography point group of the crystal should leave the physical response invariant (up to some reshuffling of indices). This imposes the constrains 
\begin{align}
\zeta^{a'b'c'} = \textrm{det}(M) M^{aa'}M^{bb'}M^{cc'} \zeta^{abc},
\label{eq:trans_pv}
\end{align}
where $\textrm{det}(M)$ introduces a negative sign for improper transformations. In the case of the observable being a vector $\pmb{\mathcal{V}}$, such as polarization, instead of a pseudovector, we would have to second order
\begin{align}
\mathcal{V}^{a}=\chi^{abc} V^{b} V^{c},
\end{align}
where $\chi$ is a third-rank tensor with transformation law
\begin{align}
\chi^{a'b'c'} =  M^{aa'}M^{bb'}M^{cc'} \chi^{abc}.
\label{eq:trans_v}
\end{align}
If the crystal point group does not contain improper transformations the factor det$(M)$ is irrelevant and the tensor and pseudotensor have the same symmetry constrains. But if the point group contains improper transformations, the symmetry constrains on tensor and pseudotensors can have profound consequences. Using Eqs.~(\ref{eq:trans_pv}) and (\ref{eq:trans_v}) we find the nonzero components of the response tensor and pseudotensor for the 32 crystallographic points groups (see Table~\ref{tab1}).

As expected, the presence of inversion symmetry forces all components of the tensor to vanish\cite{Boyd2008} while the components of the pseudotensor are finite. Even without inversion symmetry, the nonzero components are distinct for eight point groups: $-4$, 4$m $, -42$m$, 3$m$, $-6$, 6$mm$, $-6m2$, $-43m$. To see the physical implications of this we separate the symmetric and antisymmetric components of the tensor and pseudotensor (schematically),
\begin{align}
\pmb{\bar{\mathcal{V}}} &= \nu_2 |\v{V}|^2 + v_2 \v{V}\times\v{V}^{*},
\label{eq:nu_pv} \\
\pmb{\mathcal{V}} &= \sigma_2 |\v{V}|^2 + \eta_2 \v{V}\times\v{V}^{*},
\label{eq:nu_v}
\end{align}
where $\nu_2 (\sigma_2)$ is symmetric in the $V$-field indices and $v_2 (\eta_2)$ is antisymmetric in the $V$-field indices. Clearly, we can always decompose a 3rd-rank tensor (pseudo-tensor) in this way, which implies that $\zeta$ is complex with $\nu_2=\textrm{Re}(\zeta)$, $v_2=i\textrm{Im}(\zeta)$, and $(\zeta^{abc})^{*}=\zeta^{acb}$. Likewise, $\sigma_2,\eta_2$, and $\chi$ satisfy similar constrains.
\begin{table*}[]
\caption{Nonzero components of a third-rank tensor and pseudotensor for all the 32 crystallographic point groups. The external perturbation is assumed to be a vector. Also shown are  the number of symmetric ($\sigma_2,\nu_2$) and antisymmetric ($\eta_2,v_2$) distinct components in each case, see Eqs.~\ref{eq:nu_pv} and \ref{eq:nu_v}}
\begin{tabular}{|c|c|c|c|c|c|c|}
\hline
Point  &  \multicolumn{3}{c|}{tensor ~~~}            &
                                                       \multicolumn{3}{c|}{pseudo-tensor}
  \\
group  &              & $\sigma_2$ & $\eta_2$     &                                         & $\nu_2$ & $v_2$  \\
\hline
1      &  all         &   18     &   9           &  all                                    & 18      & 9     \\
\hline
-1     &  -           &   0      &   0            &  all                                    & 18     & 9     \\
\hline
\multirow{2}{*}{2}    & $xzx,xzy,xxz,xyz,yxz,yzy$     &\multirow{2}{*}{8}&\multirow{2}{*}{5}& \multirow{2}{*}{same} & \multirow{2}{*}{8} & \multirow{2}{*}{5} \\
                      & $yyz,yzx,zxx,zxy,zyx,zyy,zzz$ &                   &                   &                       &                     &                     \\
\hline
\multirow{2}{*}{$m$}  & $xzx,xzy,xxz,xyz,yxz,yzy$     &\multirow{2}{*}{8} &\multirow{2}{*}{5} & \multirow{2}{*}{same} & \multirow{2}{*}{8} & \multirow{2}{*}{5} \\
                      & $yyz,yzx,zxx,zxy,zyx,zyy,zzz$ &  &  &                       &                     &                     \\
\hline
\multirow{2}{*}{$2/m$}&  \multirow{2}{*}{-}           &\multirow{2}{*}{0}&\multirow{2}{*}{0}& $xzx,xzy,xxz,xyz,yxz,yzy$    & \multirow{2}{*}{8}&\multirow{2}{*}{5} \\
                      &                               &  &  & $yyz,yzx,zxx,zxy,zyx,zyy,zzz$&                    &                    \\
\hline
$222$                 & $xyz,xzy,yxz,yzx,zxy,zyx$     &3 &3 & same                         & 3                  & 3                  \\
\hline
$mm2$                 & $xyz,xzy,yxz,yzx,zxy,zyx$     &3 &3 & same                         & 3                  & 3                  \\
\hline
$mmm$                 &      -                        &0 &0 & $xyz,xzy,yxz,yzx,zxy,zyx$    & 3                  & 3                  \\
\hline
\multirow{2}{*}{$4$}  & $xyz=-yxz$,$xzy=-yzx$,$xzx=yzy$,$xxz=yyz$&4&3& \multirow{2}{*}{same}&\multirow{2}{*}{4} & \multirow{2}{*}{3} \\
                      &$zxx=zyy$,$zxy=-zyx$,$zzz$     &  &  &                                &                   &                    \\
\hline
\multirow{2}{*}{-$4$} & $xyz=yxz$,$xzy=yzx$,$xzx=-yzy$,$xxz=-yyz$ & 4 & 2 & $xyz=-yxz$,$xzy=-yzx$,$xzx=yzy$,$xxz=yyz$ & \multirow{2}{*}{4} & \multirow{2}{*}{3} \\
                      &  $zxx=-yzz$, $zxy=zyx$        &  &  & $zxx=zyy$, $zxy=-zyx$, $zzz$  &                   &                     \\
\hline
\multirow{2}{*}{$4/m$}& \multirow{2}{*}{-} &0&0& $xyz=-yxz$,$xzy=-yzx$,$xzx=yzy$,$xxz=yyz$& \multirow{2}{*}{4}& \multirow{2}{*}{3} \\
                      &                      & & & $zxx=zyy$,$zxy=-zyx$,$zzz$               &                   &                    \\
\hline
$422$                 & $xyz=-yxz$,$xzy=-yzx$,$zxy=-zyx$ &1&2& same                       & 1                 & 2                  \\
\hline
$4mm$   & $xzx=yzy$,$xxz=yyz$,$zxx=zyy$,$zzz$  &3 &1 & $xyz=-yxz$,$xzy=-yzx$,$zxy=-zyx$       & 1 & 2 \\
\hline
-$42m$  & $xyz=yxz$,$xzy=yzx$,$zxy=zyx$        &2&1& $xyz=-yxz$,$xzy=-yzx$,$zxy=-zyx$       & 1 & 2 \\
\hline
$4/mmm$ & -                                    &0&0& $xyz=-yxz$,$xzy=-yzx$,$zxy=-zyx$       & 1 & 2 \\
\hline
\multirow{4}{*}{$3$} & $xxx=-xyy=-yyx=-yxy$    &\multirow{4}{*}{6}&\multirow{4}{*}{3}& \multirow{4}{*}{same}&\multirow{4}{*}{6}& \multirow{4}{*}{3} \\
                     & $yyy=-xxy=-xyx=-yxx$    & & &                               &                  & \\
										 & $xxz=yyz,xzx=yzy,xyz=-yxz$   &  &  &                        &                  & \\
										 & $xzy=-yzx,zxx=zyy,zxy=-zyx,zzz$&  &  &                        &                  & \\
\hline
\multirow{4}{*}{-$3$} & \multirow{4}{*}{-} &\multirow{4}{*}{0}&\multirow{4}{*}{0}& $xxx=-xyy=-yyx=-yxy$ & \multirow{4}{*}{6} & \multirow{4}{*}{3} \\
                      &                    & & & $yyy=-xxy=-xyx=-yxx$ &                    &                    \\
											&                    & & & $xxz=yyz$, $xzx=yzy$, $xyz=-yxz$  &        &                    \\
											&                    & & & $xzy=-yzx$ $zxx=zyy$,$zxy=-zyx$, $zzz$ &  &                     \\
\hline
\multirow{2}{*}{$32$} &$xxx=-xyy=-yyx=-yxy$&2&2                  &\multirow{2}{*}{same}& \multirow{2}{*}{2}&\multirow{2}{*}{2} \\
                      &$xyz=-yxz$, $xzy=-yzx$, $zxy=-zyx$ & &    &                     &                   &                   \\
\hline
\multirow{2}{*}{$3m$} &$yyy=-xxy=-xyx=-yxx$&4&1                  & $xxx=-xyy=-yyx=-yxy$& \multirow{2}{*}{2}& \multirow{2}{*}{2} \\
                      & $xzx=yzy$,$xxz=yyz$,$zxx=zyy$,$zzz$& &   & $xyz=-yxz$, $xzy=-yzx$,$ zxy=-zyx$ &    &                    \\
\hline
\multirow{2}{*}{-$3m$}& \multirow{2}{*}{-}                 &0&0  & $xxx=-xyy=-yyx=-yxy$ & \multirow{2}{*}{2} & \multirow{2}{*}{2} \\
                      &                                    & &   & $xyz=-yxz$, $xzy=-yzx$,$ zxy=-zyx$ &      &                    \\
\hline
\multirow{2}{*}{$6$} & $xxz=yyz$, $xyz=-yxz$, $xzx=yzy$,   &4&3  & \multirow{2}{*}{same}& \multirow{2}{*}{4} & \multirow{2}{*}{3} \\
                     & $xzy=-yzx$, $zxx=zyy$, $zxy=-zyx$, $zzz$&&&                      &                    & \\
\hline
\multirow{2}{*}{-$6$}& $xxx=-xyy=-yyx=-yxy$  &2 &0      & $xxz=yyz$,$xyz=-yxz$,$xzx=yzy$ & \multirow{2}{*}{4} & \multirow{2}{*}{3} \\
                     & $yyy=-xxy=-xyx=-yxx$  &  &      & $xzy=-yzx$,$zxx=zyy$,$zxy=-zyx$,$zzz$ &             &                    \\
\hline
\multirow{2}{*}{$6/m$} & \multirow{2}{*}{-}  &0 &0     & $xxz=yyz$,$xyz=-yxz$,$xzx=yzy$  & \multirow{2}{*}{4} & \multirow{2}{*}{3} \\
                       &                     &  &      & $xzy=-yzx$,$zxx=zyy$,$zxy=-zyx$,$zzz$ &              &                    \\
\hline
$622$   & $xyz=-yxz$, $xzy=-yzx$, $zxy=-zyx$ &1 &2     & same                             & 1 & 2              \\
\hline
$6mm$   & $xzx=yzy$,$xxz=yyz$,$zxx=zyy$,$zzz$&3 &1     & $xyz=-yxz$,$xzy=-yzx$,$zxy=-zyx$  & 1 & 2              \\
\hline
-$6m2$  & $yyy=-yxx=-xxy=-xyx$               &1 &0     &  $xyz=-yxz$,$xzy=-yzx$,$zxy=-zyx$ & 1 & 2              \\
\hline
$6/mmm$ &  -                                 &0 &0     &  $xyz=-yxz$,$xzy=-yzx$,$zxy=-zyx$ & 1 & 2              \\
\hline
$23$    & $xyz=yzx=zxy$,$xzy=yxz=zyx$        &1 &1     &  same                             & 1 & 1              \\
\hline
$m$-$3$ &  -                                 &0 &0     & $xyz=yzx=zxy$,$xzy=yxz=zyx$       & 1 & 1              \\
\hline
$432$   & $xyz=yzx=zxy=-xzy=-yxz=-zyx$       &0 &1     & same                              & 0 & 1              \\
\hline
-$43m$  & $xyz=yzx=zxy=xzy=yxz=zyx$          &1 &0     & $xyz=yzx=zxy=-xzy=-yxz=-zyx$      & 0 & 1              \\
\hline
$m$-$3m$& -                                  &0 &0     & $xyz=yzx=zxy=-xzy=-yxz=-zyx$      & 0 & 1              \\
\hline
\end{tabular}
\label{tab1}
\end{table*}
\begin{table*}[]
\caption{Symmetric $\sigma_2$ ($\nu_2$) and antisymmetric $\eta_2$ ($v_2$) distinct nonzero components of a third-rank tensor (pseudotensor).}
\begin{center}
\begin{tabular}{|c|c|c|c|c|}
\hline
Point  &  \multicolumn{2}{c|}{tensor}         &     \multicolumn{2}{c|}{pseudo-tensor}                      \\
group  &     $\sigma_2$ & $\eta_2$             & $\nu_2$ & $v_2$   \\
\hline
\multirow{3}{*}{1}  &  $xxx,xyy,xzz,yxx,yyy,yzz$     &   $xxy,xxz,xyz,yxy$  &  \multirow{3}{*}{same}  & \multirow{3}{*}{same}     \\
                    &  $zxx,zyy,zzz,xxy,xxz,xyz$     &   $yxz,yyz,zxy,zxz$  &         &    \\
										&  $yxy,yxz,yyz,zxy,zxz,zyz$     &   $zyz$              &         &     \\
\hline
\multirow{3}{*}{-1} &  \multirow{3}{*}{-}  & \multirow{3}{*}{-} & $xxx,xyy,xzz,yxx,yyy,yzz$ & $xxy,xxz,xyz,yxy$   \\
                    &                      &                    & $zxx,zyy,zzz,xxy,xxz,xyz$ & $yxz,yyz,zxy,zxz$    \\
                    &                      &                    & $yxy,yxz,yyz,zxy,zxz,zyz$ & $zyz$ \\
\hline
\multirow{2}{*}{2} &  $xzx,xzy,yxz,yyz$  & $xzx,xzy,yxz$ & \multirow{2}{*}{same} & \multirow{2}{*}{same}   \\
                   &  $zxx,zxy,zyy,zzz$  & $yyz,zxy$             &                       &     \\
\hline
\multirow{2}{*}{$m$}&  $xzx,xzy,yxz,yyz$  & $xzx,xzy,yxz$ & \multirow{2}{*}{same} & \multirow{2}{*}{same}   \\
                    &  $zxx,zxy,zyy,zzz$  & $yyz,zxy$             &                       &     \\
\hline
\multirow{2}{*}{$2/m$}&  \multirow{2}{*}{-}  & \multirow{2}{*}{-} & $xzx,xzy,yxz,yyz$ & $xzx,xzy,yxz$   \\
                      &                      &                    & $zxx,zxy,zyy,zzz$ &  $yyz,zxy$   \\
\hline
$222$ & $xyz,yxz,zxy$ & $xyz,yxz,zxy$ & same & same  \\
\hline
$mm2$ & $xyz,yxz,zxy$ & $xyz,yxz,zxy$ & same & same  \\
\hline
$mmm$ & - & - & $xyz,yxz,zxy$  & $xyz,yxz,zxy$  \\
\hline
\multirow{2}{*}{$4$}&  $xyz=-yxz,xzx=yzy$  & $xyz=-yxz,xzx=yzy$ & \multirow{2}{*}{same} & \multirow{2}{*}{same}   \\
                    &  $zxx=zyy,zzz$       & $zxy$              &                       &     \\
\hline
\multirow{2}{*}{$-4$}&  $xyz=yxz,xzx=-yzy$  & $xyz=yxz$ & $xyz=-yxz,xzx=yzy$ & $xyz=-yxz,xzx=yzy$   \\
                     &  $zxx=-yzz,zxy=zyx$  & $xzx=-yzy$& $zxx=zyy,zzz$      & $zxy$      \\
\hline
\multirow{2}{*}{$4/m$}& \multirow{2}{*}{-} & \multirow{2}{*}{-} & $xyz=-yxz,xzx=yzy$ & $xyz=-yxz,xzx=yzy$   \\
                      &   &   & $zxx=zyy,zzz$      & $zxy$      \\
\hline
422   & $xyz=-yxz$  & $xyz=-yxz,zxy$ & same & same   \\
\hline
\multirow{2}{*}{$4mm$}&  $xzx=yzx,zxx=zyy$ & \multirow{2}{*}{$xzx=yzy$}& \multirow{2}{*}{$xyz=-yxz$} & $xyz=-yxz$   \\
                      &  $zzz$             &                           &                             & $zxy$      \\
\hline
$-42m$& $xyz=yxz,zxy$ & $xyz=yxz$ & $xyz=-yxz$ & $xyz=-yxz,zxy$   \\
\hline
$4/mmm$& - & - & $xyz=-yxz$ & $xyz=-yxz,zxy$   \\
\hline
\multirow{4}{*}{3} & $xxx=-xyy=-yyx$ &                   & \multirow{4}{*}{same} & \multirow{4}{*}{same}   \\
                   & $yyy=-xxy=-xyx$ & $xxz=yyz,xyz=-yxz$&                       &     \\
                   & $xxz=yyz,xyz=-yxz$   & $zxy$             &                       &  \\
                   & $zxx=zyy,zzz$        &                   &                       &  \\
\hline
\multirow{4}{*}{-3}& \multirow{4}{*}{-}&\multirow{4}{*}{-}& $xxx=-xyy=-yyx$    &    \\
                   &                   &                  & $yyy=-xxy=-xyx$    &  $xxz=yyz,xyz=-yxz$   \\
                   &                   &                  & $xxz=yyz,xyz=-yxz$ &  $zxy$ \\
                   &                   &                  & $zxx=zyy,zzz$      &  \\
\hline
\multirow{2}{*}{$4mm$}&  $xzx=yzx,zxx=zyy$ & \multirow{2}{*}{$xzx=yzy$}& \multirow{2}{*}{$xyz=-yxz$} & $xyz=-yxz$   \\
                      &  $zzz$             &                           &                             & $zxy$      \\
\hline
\multirow{2}{*}{$32$}&  $xxx=-xyy=-yyx$ & $xyz=-yxz$ & \multirow{2}{*}{same} & \multirow{2}{*}{same}   \\
                     &  $xyz=-yxz$      & $xzy$      &                       &       \\
\hline
\multirow{2}{*}{3$m$}& $yyy=-xxy=-yxx$   & \multirow{2}{*}{$xzx=yzy$}& $xxx=-xyy=-yyx$ & $xyz=-yxz$   \\
                     & $xzx=yzy,zxx=zyy,zzz$ &                       & $xyz=-yxz$      & $zxy$    \\
\hline
\multirow{2}{*}{-3$m$}& \multirow{2}{*}{-}& \multirow{2}{*}{-}& $xxx=-xyy=-yyx$ & $xyz=-yxz$   \\
                      &                   &                   & $xyz=-yxz$      & $zxy$    \\
\hline
\multirow{2}{*}{6}& $xxz=yyz,xyz=-yxz$ & $xxz=yyz,xyz=-yxz$ & \multirow{2}{*}{same} & \multirow{2}{*}{same}   \\
                  & $zxx=zyy,zzz$      & $zxy$              &       &     \\
\hline
\multirow{2}{*}{-6}& $xxx=-xyy=-yyx$ & \multirow{2}{*}{-} & $xxz=yyz,xyz=-yxz$& $xxz=yyz,xyz=-yxz$   \\
                   & $yyy=-xxy=-yxx$ &                    &  $zxx=zyy,zzz$    & $zxy$    \\
\hline
\multirow{2}{*}{6/$m$}& \multirow{2}{*}{-} & \multirow{2}{*}{-} & $xxz=yyz,xyz=-yxz$& $xxz=yyz,xyz=-yxz$   \\
                      &                    &                    & $zxx=zyy,zzz$     & $zxy$    \\
\hline
622   & $xyz=-yxz$ & $xyz=-yxz,zxy$ & same & same   \\
\hline
6$mm$ & $xzx=yzy,zxx=zyy,zzz$ & $xzx=yzy$ & $xyz=-yxz$ & $xyz=-yxz,zxy$   \\
\hline
-6$m2$ & $yyy=-yxx=-xxy$ & - & $xyz=-yxz$ & $xyz=-yxz,zxy$   \\
\hline
Point  &  \multicolumn{2}{c|}{tensor}         &     \multicolumn{2}{c|}{pseudo-tensor}                      \\
group  &     $\sigma_2$ & $\eta_2$             & $\nu_2$ & $v_2$   \\
\hline
6$/mmm$ & - & - & $xyz=-yxz$ & $xyz=-yxz,zxy$   \\
\hline
23 & $xyz=yzx=zxy$ & $xyz=yzx=zxy$ & same & same   \\
\hline
$m$-3 & - & - & $xyz=yzx=zxy$ & $xyz=yzx=zxy$    \\
\hline
432 & - & $xyz=yzx=zxy$ & same & same   \\
\hline
-43$m$ & $xyz=yzx=zxy$ & - & - & $xyz=yzx=zxy$   \\
\hline
$m$-$3m$ & - & - & - & $xyz=yzx=zxy$   \\
\hline
\end{tabular}
\end{center}
\label{tab:nu2_v2}
\end{table*}
The distinct nonzero components of $\nu_2, v_2$ ($\sigma_2,\eta_2$) can be found simply by inspection (see Table~\ref{tab:nu2_v2}). Physically, $\nu_2$ ($\sigma_2$) determines the response to LPE whereas $v_2$ ($\eta_2$) determines the response to CPE; i.e., the former vanishes for CPE and the latter vanishes for LPE. This decomposition provides information about whether a material supports spin polarization, second-harmonic generation, current, etc., via CPE, LPE, CPE or LPE, or neither. 

For example, Table~\ref{tab1} shows that \textit{all} 32 points groups admit spin polarization via CPE (usual spin orientation/inverse spin Faraday effect), but only 29 admit it with LPE. The points groups $m$-3$m$, -43$m$, and $432$ are too symmetric and all $\nu_2$  components vanish. In other words, there is no material whose spin response is pure LPE. This is one of our main results. There are three (and only three) points groups whose tensor response is purely LPE active (-43$m$, -6$m$2, -6) which means, e.g., only a shift current\cite{Sturman1992,Baltz1981,Sipe2000,Fregoso2019,IbanezAzpiroz2018,Holder2021,Ma2021,Ahn2022,Zhu2024,Jankowski2023,Resta2024} can be generated in these materials. GaAs (point group -43$m$) is an example which is known to admit shift current\cite{Nastos2006} but not an injection current. Similarly GaAs exhibits second-harmonic generation only for LPE. There is only one point group (432) whose tensor response is purely CPE active. Depending on the application (e.g., solar cell, optoelectronic switch, etc.) one may want a material which responds to one (or two) particular stimuli. The general scheme we present here is of help in guiding this choice.


\section{Material examples}
Using Density Functional Theory (DFT) we compute the spectrum of the spin response pseudotensor as a function of light frequency for nonmagnetic inversion-symmetric SnS$_2$ and Si and compare with that of inversion-asymmetric Te, GaAs and InSb. The equations for the sprin response pseudotensor in terms of Bloch states are given in the Appendix~\ref{sec:bpse_ins}.

\subsection{Te (point group $32$)}
Te has spin-split bands and is widely used in technological applications. A decomposition of the spin response pseudotensor into its symmetric and antisymmetric components gives two distinct symmetric components $\nu_2^{xxx}=-\nu_2^{xyy}=-\nu_2^{yyx}$, and $\nu_2^{xyz}=-\nu_2^{yzx}$ and two distinct antisymmetric components $v_2^{xyz}=-v_2^{yxz}$, and $v_2^{zxy}$. 

\begin{figure}[t!]
\subfigure{\includegraphics[width=.47\textwidth]{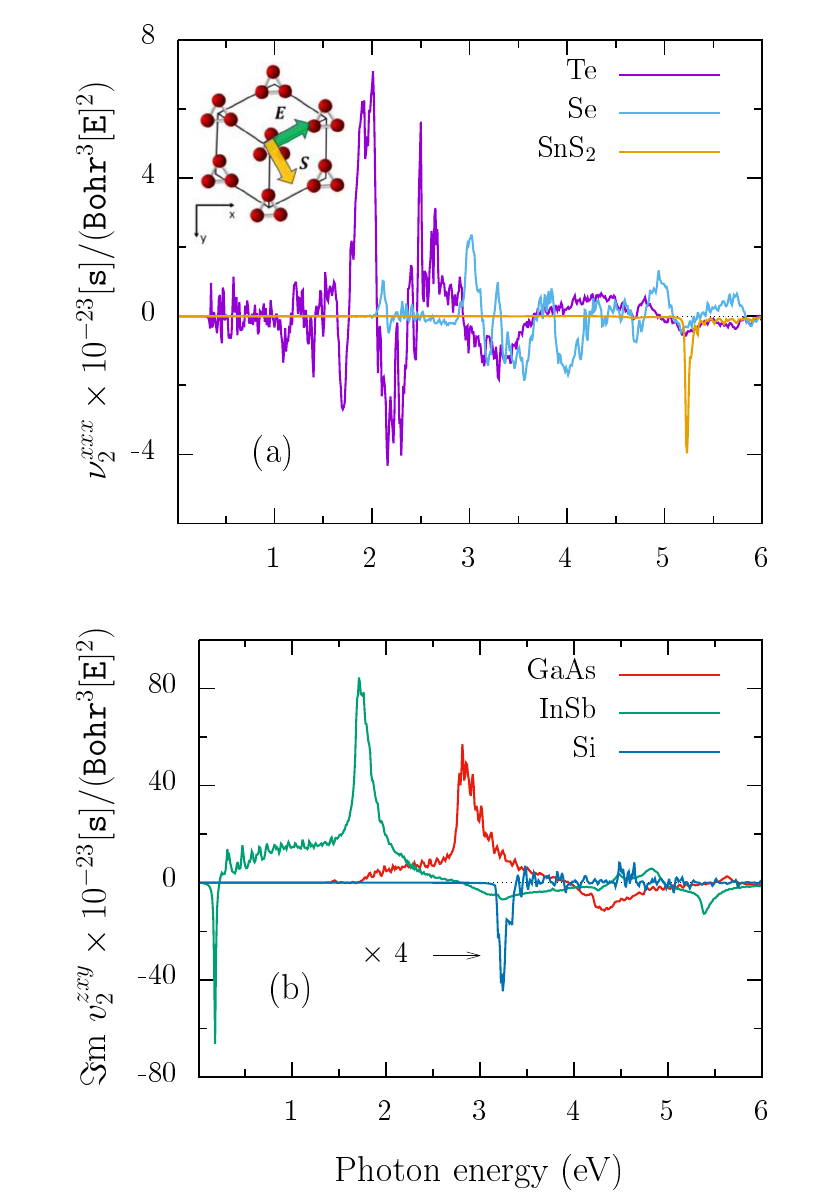}}
\caption{(a) Spectrum of $\nu_{2}^{xxx}$ of Te, Se, and SnS$_2$. (b) Spectrum of $v_2^{zxy}$ of GaAs, InSb, and Si calculated using DFT (see Table~\ref{tab3}).  Te, Se and SnS$_2$ admit LPE spin polarization because some $\nu_2$ are finite. GaAs and InSb, on the other hand, only admit CPE spin polarization. SnS$_2$ has double spin-degenerate bands and yet has a large spin response at $5.2$ eV. The units of the response pseudotesor are $(\hbar/2)/a_0^3 (V/m)^2$ where $a_0$ is the Bohr radius.}
\label{fig:nu_xxx_v_zxy}
\end{figure}

Let us consider an electric field with linear polarization in the $xy$-plane given by $\v{E}=(\cos{\phi},\sin{\phi},0) E_0 \cos{\omega t}$. $\phi$ is the angle of the field with respect to $x$-axis and $E_0$ is the electric field amplitude [see inset of Fig.~\ref{fig:nu_xxx_v_zxy}(a)]. The induced spin is then 
\begin{align}
\v{S}^{\leftrightarrow}= (\cos{2\phi},-\sin{2\phi},0)S_0^{\leftrightarrow},
\label{eqn:S_Te}
\end{align}
which also lies in the $xy$-plane and where $S_0^{\leftrightarrow}= E_0^2 \nu_{2}^{xxx}/2$ is the spin amplitude. A rough estimate of the spin magnetization is then
\begin{align}
M_0^{ \leftrightarrow}=\frac{2\mu_{B} S_0^{\leftrightarrow}}{\hbar},
\label{eq:Mag_eq}
\end{align}
where $\mu_B=9.27\times 10^{-24}$ J/T is the Bohr magneton. Note that the induced spin is parallel to the electric field when $\phi=0,\pi/3,2\pi/3,\pi$ but it is perpendicular when $\phi= \pi/6, 3\pi/6, 5\pi/6, 7\pi/6, 9\pi/6, 11\pi/6$. As can be seen, the direction of the in-plane electric field provides a fine control over the induced spin. This should be compared and contrasted with only two distinct options available using circularly polarized light: right/left circular polarization.

In Fig.~\ref{fig:nu_xxx_v_zxy}(a) we show the spectrum of $\nu_2^{xxx}$ for Te, Se and SnS$_2$ computed from DFT. These calculations include spin-orbit coupling via a norm-conserving relativistic separable dual space Gaussian pseudopotentials of Hartwigsen-Goedecker-Hutter\cite{Hartwigsen1998} and the modified Becke-Jonhson (metaGGA) functional\cite{Rasanen2010}. To account for the under estimation of the band gap, we applied the scissors operator with scissor shifts of 0.1, 0.236 and 0.734 eV for Te, Se, and SnS$_2$, respectively. The convergence of the spectra was reached at the cutoff energy of $E_c= 25$ Ha; 58 conduction bands for Te and Se, and 34 for SnS$_2$; 15062, 14256 and 14525 $\v{k}$-points for Te, Se and SnS$_2$, respectively. In Table~\ref{tab3} we show the relaxed structures and band gaps used in the DFT calculation.

From Fig.~\ref{fig:nu_xxx_v_zxy}(a) we read off the maximum value of $\nu_2^{xxx} \sim 6\times 10^{-23}$ $(\hbar/2)/a_0^3\times$(V/m)$^2$ at about $\hbar\omega=2$ eV. Then using Eqn.~(\ref{eq:Mag_eq}) we estimate the magnitude of the maximum spin magnetization. As example, for these DFT calculations we use a typical relaxation time in solids $\hbar/\tau = 0.01$ eV (or 60 ps) and an attainable electric field amplitude of $E_0=10^9$ m/V.  We find that Te can achieve a spin magnetization of about $\mu_0 M=24$ Gauss which is comparable to  naturally occuring Ferromagnets as seen in Table.~\ref{tab:magnetic_mom}. 

\begin{table}[]
\caption{Comparison of spontaneous magnetization of some typical ferromagnets~\cite{Ashcroft1976} versus photomagnetization. Indicated in bold are the peak photomagnetizations obtained from DFT calculations. The magnetization is given in units of $\mu_0 M_0$ where $\mu_0$ is the permittivity of free space and $M_0$ is the peak magnetization. The polarizations of light are indicated as $\circlearrowright$ = circular, $\leftrightarrow$ = linear.}
\begin{tabular}{c|c}
\hline
\hline
\multirow{2}{*}{Material} &  Magnetic moment      \\
                          &  ($\mu_0$M$_0$ in Gauss)    \\
\hline
Fe               & 1752  ~~~~\\
Ni               &  510  ~~~\\
InSb             &\textcolor{red}{315} $\circlearrowright$ \\
CrBr$_3$         & 270 ~~~\\
GaAs             &\textcolor{red}{197} $\circlearrowright$ \\
CuFe$_2$O$_3$    & 160 ~~~~\\
Si               &\textcolor{red}{39} $\circlearrowright$ \\ 
Te               &\textcolor{red}{24} $\leftrightarrow$ \\
SnS$_2$          &\textcolor{red}{16} $\leftrightarrow$ \\ 
\hline
\end{tabular}
\label{tab:magnetic_mom}
\end{table}


\subsection{SnS$_2$ (point group -3$m$)}
SnS$_2$ has the same trigonal crystal system as Te and Se and the same nonzero $\nu_2$ and $v_2$ components (see Table \ref{tab3}). SnS$_2$ however, has inversion symmetry and hence all quadratic (vector) responses such as current and second-harmonic generation vanish. Inversion and time-reversal symmetric materials such as SnS$_2$ have double degenerate spin bands and would not be expected to spin-polarize with LPE. Yet, if we consider an electric field is given by $\v{E}=(\cos{\phi},\sin{\phi},0) E_0 \cos{\omega t}$, where $\phi$ is the angle of the field with respect to $x$ axis and $E_0$ is the electric field amplitude, we find that the induced spin is also given by Eq.~(\ref{eqn:S_Te}). Fig.~\ref{fig:nu_xxx_v_zxy}(a) shows the spectrum of SnS$_2$. In general, the pseudotensor for SnS$_2$ is smaller than that of Te, but at $5.2$ eV it reaches $\nu_2^{xxx} \sim -4\times 10^{-23}$ $(\hbar/2)/a_0^3\times$(V/m)$^2$ comparable to Te. The maximum magnetization is 16 G. 

%

\subsection{GaAs, InSb (point group \- -43$m$) and Si (point group $m$-3$m$)}
GaAs and InSb (point group -43$m$) and Si (point group $m$-3$m$) have zinc blend and diamond structures, respectively, with one distinct antisymmetric component $v_2^{xyz}=v_2^{yzx}=v_2^{zxy}$ and no symmetric components, i.e., they exhibit only CPE spin polarization. Since Si has inversion symmetry (small SOC) we would expect a stronger polarization effect in GaAs and InSb compared with Si. Also, since InSb has a larger SOC than GaAs we would expect stronger spin polarization in InSb than in GaAs. Indeed this is what we see in our DFT computations [see Fig.~\ref{fig:nu_xxx_v_zxy}(b)]. GaAs, InSb and Si spin responses reach convergence at $E_c=$20; Ha  22,  34, and 136 conduction bands; and  75671 and 7464 $k$ points, correspondingly. The used scissor shifts were 0.289, 0.112 and 0.322 eV for GaAs, InSb, and Si, respectively. 

If we consider an electric field circularly polarized in the $xy$-plane given by $\v{E}=E_0(\cos{\omega t},\pm\sin{\omega t},0)$ the induced spin is perpendicular to the $xy$-plane and given by
\begin{align}
\v{S}^{\circlearrowleft}=\mp \hat{\v{z}}S_0^{\circlearrowleft},
\label{eq:GaAs_S}
\end{align}
where $S_0^{\circlearrowleft}=i v_2^{zxy} E_0^2$ is the spin amplitude. In Fig.~\ref{fig:nu_xxx_v_zxy}(b) we see the spectrum of $v_2^{zxy}$. The maximum magnetization for InSb is the highest (315 G), the second highest is for GaAs (197 G), followed by Si (39 G), in agreement with our intuition (see Table~\ref{tab:magnetic_mom}).

\begin{table*}[]
\caption{DFT parameters and maximum spin photomagnetization of prototypical semiconductors with and without inversion symmetry. Nonzero $\nu_2$ and $v_2$ components, equilibrium lattice constants, and direct band gaps are indicated. The peak spin magnetization $M_0$ (in units of $\mu_0 M_0$) is estimated from the largest response using Eq.~(\ref{eq:Mag_eq}) or Eq.~(\ref{eq:GaAs_S}). We use attainable $E_0=10^{9}$ V$/$m and $\hbar/\tau=0.01$ eV. $\mathcal{I}(\mathcal{T})$ indicates inversion (time reversal) symmetry.}
\begin{tabular}{c|c|c|c|c|c|c|c|c|c}	
\hline
\hline
\multirow{3}{*}{} & & \multirow{3}{*}{$\mathcal{I}$}  & \multirow{3}{*}{$\mathcal{T}$} & \multirow{3}{*}{$\nu_2$}  & \multirow{3}{*}{$v_2$}   & Lattice constants &  Band gap (eV)  & &   \\
		  &  Point    &      &  &       &        & (\AA) & DFT ~~~~~~~~~~~~~~~~& $\mu_0 M_{0}$   & $\hbar\omega_{0}$    \\
			&  group   &    &      & &   &  $a$~~~~~~~~$b$~~~~~~~~$c$  & {\small metaGGA} ~~~~~~~~ Exp.~ & Gauss  & (eV)  \\
\hline
InSb & -43$m$ & \ding{55}& \checkmark&- & $xyz=yzx=zxy$ & \multirow{1}{*}{6.48~~~~6.48~~~~6.48}\cite{Vurgaftman2001}&\multirow{1}{*}{0.07 (i)~~~~~~~~~~0.18}\cite{Adachi1999} & 315 &  1.7  \\
\hline
GaAs & -43$m$&  \ding{55}  &\checkmark & - & $xyz=yzx=zxy$  & \multirow{1}{*}{5.65~~~~5.65~~~~5.65}\cite{Vurgaftman2001} & \multirow{1}{*}{1.13 (d)~~~~~1.42-1.43}\cite{Vurgaftman2001} &  197 &  2.8  \\ 
\hline
Si & $m$-3$m$ & \checkmark& \checkmark& -  &  $xyz=yzx=zxy$  & \multirow{1}{*}{5.43~~~~5.43~~~~5.43}\cite{Sze2007}&\multirow{1}{*}{0.798
(i)~~~~~~~~~~1.12}\cite{Sze2007}& 39 &   3.2  \\
\hline
\multirow{2}{*}{Te}   &  \multirow{2}{*}{32}   & \multirow{2}{*}{\ding{55}} & \multirow{2}{*}{\checkmark} & $xxx=-xyy=-yyx$ &  $xyz=yzx$ & \multirow{2}{*}{4.46~~~~4.46~~~~5.93\cite{Orlov2022}} & 0.22 (d)~~~~ 0.32 \cite{Orlov2022}& \multirow{2}{*}{24}& \multirow{2}{*}{2.0}   \\
 & & & & $xyz=-yzx$ & $zxy$ & &    &  & \\
\hline
\multirow{2}{*}{SnS$_2$}  & \multirow{2}{*}{-3$m$}  & \multirow{2}{*}{\checkmark} & \multirow{2}{*}{\checkmark}& $xxx=-xyy=-yyx$ &  $xyz=yzx$  & \multirow{2}{*}{3.70~~~~3.70~~~~6.98}~~~~~ & \multirow{2}{*}{1.75 (i)~~~~~~~~~~~2.48\cite{Burton2016}}& \multirow{2}{*}{16} & \multirow{2}{*}{5.2} \\ 
& & & & $xyz=-yzx$ & $zxy$ & & & &\\
\hline
\hline
\end{tabular}
\label{tab3}
\end{table*}

\section{Discussion}
We showed that all 32 crystallography point groups admit CPE spin polarization, but only 29 of those admit LPE spin polarization. The point groups 432,-43$m$ and $m$-$3m$ do not develop LPE-spin polarization. As shown in Ref.~\cite{Fregoso2022}, CPE spin polarization does not require SOC, inversion symmetry or quantum coherence. LPE spin polarization, on the other hand, requires SOC, quantum coherence, and, as shown here, a crystal that is not too symmetric. An intuitive picture emerges as follows: at every $\v{k}$ point in the Brillouin zone (BZ) there is a slightly different local magnetic field (due to SOC), and hence spin-up and spin-down states precess about slightly different directions. Addition of all such spins gives a net polarization only for certain types trajectories in the BZ. For highly symmetric crystals the addition of spins cancels. Because the off-diagonal components of the spin are involved in the sum we require quantum coherence. 

For practical purposes, LPE spin polarization provides a finer control knob over the spin, namely, the electric field polarization direction. We have shown with realistic materials' DFT computations that photomagnetization with LPE and CPE could be comparable (Table~\ref{tab:magnetic_mom}) hence broadening options for technological applications in magnet-optics. We have not discussed the contribution to the orbital magnetization which we expect to be present. In a future work we will address this important topic. Our results also give more insight into the materials studied recently in Ref.~\cite{Ofer2023}. For example $BiH$ has point point -3$m$ and hence allows LPE(CPE) spin polarization with two components $xxx=-xyy=-yyx, xyz=-yxz$  ($xyz=-yxz, zxy$).

Because of the requirement of quantum coherence we expect our results to be observed at low enough temperatures and ultrashort time scales, e.g., pump-probe experiments. Note also that the classification of responses using spatial groups (instead of magnetic groups) assumes weak SOC.

\section{Acknowledgments}
B.S.M. acknowledgeS support from a sabbatical fellowship and the hospitality of the Max Planck Institute for the Structure and Dynamics of Matter in Hamburg, Germany. B.M.F. acknowledges support from NSF grant nO. DMR-2015639.

\appendix

\section{Theory of LPE and CPE spin polarization}
\label{sec:bpse_ins}
We consider an insulator with fully occupied valence bands and fully empty conduction bands. Let us assume there is a monochromatic optical field of the form $\v{E} = \v{E}(\omega) e^{-i\omega t}+ \mathrm{c.c.}$  The \textit{static} induced spin to second order in the electric field is
\begin{align}
S^{a}= 2 \zeta^{abc}(0;\omega,-\omega) E^{b}(\omega) E^{c}(-\omega).
\label{S2ndnFSexample}
\end{align}
We now separate the symmetric and antisymmetric responses of the \textit{interband} response by defining
\begin{align}
\nu_{2}^{abc} &\equiv (\zeta^{abc} + \zeta^{acb})/2,
\label{eq:nu_upsilon_defs1} \\
\upsilon_{2}^{abc} &\equiv (\zeta^{abc} - \zeta^{acb})/2.
\label{eq:nu_upsilon_defs2}
\end{align}
Equation (\ref{S2ndnFSexample}) then becomes
\begin{align}
S^{(2)a}= 2 \nu_{2}^{abc}& E^{b}(\omega) E^{c}(-\omega)+ 2 \upsilon_{2}^{abc} E^{b}(\omega) E^{c}(-\omega),
\label{S2ndnFSexample2}
\end{align}
which is of the form of Eq.~(\ref{eq:nu_pv}). The tensor $\zeta^{abc}$ can be further decomposed into 2-band and 3-band contributions,\cite{Fregoso2022}
\begin{align}
\zeta= \zeta_{2b} +\zeta_{3b},
\end{align}
where
\begin{align}
\zeta_{2b}^{abc} &\hspace{-2pt}= \hspace{-2pt}-\frac{ie^2}{2\hbar^2 V} \hspace{-2pt} \sum_{nm\v{k}} \frac{s_{nm}^a }{\bar{\omega}_{nm}} \bigg[ \hspace{-2pt} \left(\frac{r_{mn}^b f_{nm}}{\omega_{mn}-\bar{\omega}}\right)_{;c} \hspace{-3pt}+\hspace{-3pt} \left(\frac{r_{mn}^c f_{nm}}{\omega_{mn}+\bar{\omega}^{*}}\right)_{;b} \hspace{-2pt}\bigg],
\label{eq:zeta2nFS2b} \\
\zeta_{3b}^{abc} &= \frac{e^2}{2\hbar^2 V}\sum_{nml\v{k}} \frac{s_{nm}^a}{\bar{\omega}_{nm}} \bigg[\frac{r_{ml}^b r_{ln}^c f_{lm}}{\omega_{ml}-\bar{\omega}} - \frac{r_{ml}^c r_{ln}^b f_{nl}}{\omega_{ln}-\bar{\omega}} \nn \\
&~~~~~~~~~~~~~~~~~~~~~~+ \frac{r_{ml}^c r_{ln}^b f_{lm}}{\omega_{ml}+\bar{\omega}^{*}} - \frac{r_{ml}^b r_{ln}^c f_{nl}}{\omega_{ln}+\bar{\omega}^{*}}\bigg],
\label{eq:zeta2nFS3b}
\end{align}
and $\bar{\omega} \equiv \omega + i/\tau$ and $\bar{\omega}_{nm} = \omega_{nm} + i/\tau$. The notation used is the same as in \cite{Fregoso2022}. Although not obvious, $\nu_2$ depends on the off-diagonal elements of density matrix whereas $v_2$ depends on the diagonal elements of the density matrix. This means $\nu_2$ processes (such as LPE spin polarization) require quantum coherence. It is easy to check that $(\zeta^{abc})^{*}=\zeta^{acb}$ and hence
\begin{align}
\nu_{2} & = \textrm{Re}[\zeta],
\label{eq:nu_upsilon_defs12} \\
\upsilon_{2} & = i\textrm{Im}[\zeta].
\label{eq:nu_upsilon_defs22}
\end{align}
These equations are used within DFT to compute $\nu_2$ and $v_2$.


%

\end{document}